# The center of the Type Ia supernova remnant SNR 0509-67.5 is empty of any ex-companion star to $M_V$=+8.4

Bradley E. Schaefer[1] & Ashley Pagnotta[1]

[1]Department of Physics and Astronomy, Louisiana State University, Baton Rouge, Louisiana 70803, USA.

**Type Ia supernova (SNe Ia) are thought to originate in the explosion of a white dwarf[1]. The explosion could be triggered by the merger of two white dwarfs[2,3] ('double-degenerate' origin), or by mass transfer from a companion star[4,5] (the 'single-degenerate' path). The identity of the progenitor is still controversial; for example, a recent argument against the single-degenerate origin[6] has been widely rejected[7-11]. One way to distinguish between the double- and single-degenerate progenitors is to look at the center of a known SN Ia remnant to see whether any former companion star is present[12,13]. A likely ex-companion star for the progenitor of Tycho's supernova has been identified[14], but that claim is still controversial[15-18]. Here we report that the central region of the supernova remnant SNR 0509-67.5 (the site of a Type Ia supernova 400±50 years ago, based on its light echo[19,20]) in the Large Magellanic Cloud contains no ex-companion star to a limit of $V$=26.9 magnitude ($M_V$=+8.4) within the extreme 99.73% region with radius 1.43". The lack of any ex-companion star to deep limits rules out all published single-degenerate models. The only remaining possibility is that the progenitor for this particular SN Ia was a double-degenerate system.**

The progenitor of any SN Ia has never been identified. Various candidate classes have been proposed (see Table 1 and Supplementary Information section 1), although arguments and counterarguments have resulted in no decisive solution. It is possible that the observed SNe Ia might have two comparable-sized progenitor classes[21]. In double-degenerate systems, the two white dwarfs will both be completely destroyed by the supernova explosion. In single-degenerate systems, the mass-donor star (orbiting the doomed white dwarf) will survive the explosion, and shine at near its pre-explosion brightness from the middle of the expanding supernova remnant. (During the explosion, portions of the outer envelope of the companion star will be stripped off[22,23], but its location on the color-magnitude diagram will not change greatly[24].) The program of distinguishing between the progenitor models by looking for an ex-companion star inside a known SN Ia remnant has been attempted only once[14], for Tycho's supernova of 1572. A particular G-type subgiant star has been identified as being the ex-companion, and if so, it would point to a recurrent nova as the progenitor for Tycho's supernova[14]. Several concerns have been raised[15,17] concerning this identification and these have been answered[18], although the case remains unresolved.

To break this impasse, we look to a supernova remnant in the nearest galaxy to our own, the Large Magellanic Cloud (LMC), as we consider the case of SNR 0509-67.5, which was an SN Ia (of the SN1991T class) 400±50 years ago[19,20,25,26]. SNR 0509-67.5 has excellent public domain images that were taken by the *Hubble Space Telescope* (*HST*). All of the stars in the field have been measured for B, V, and I magnitude with standard *IRAF* aperture photometry and set to Vega magnitudes with the standard calibration (see Table 2). The faintest visible star (at the 5-σ detection level) is at $V$=26.9.

If any ex-companion still exists after the explosion ~400 years ago, then it must be located near the center of the remnant. We have measured the geometric center of the shell with three independent methods (see Supplementary Information section 2): using the edge of the Hα shell, the edge of the X-ray shell, and the minimum of the Hα light in the interior of the remnant. Each of these three derived centers are from different gas and regions, so they are independent and provide a measure of the statistical and systematic uncertainties in the center position. Our combined geometric center is at J2000 05:09:31.208, -67:31:17.48, with 1-σ uncertainties of 0.14" along the short axis (roughly ENE to WSW) and 0.20" along the long axis (tilted 18°±3° to the west of north).

The position of any ex-companion star will be offset from the estimated geometric center of the shell due to measurement errors of the center position, proper motion of the star, and asymmetries in the shell. The proper motion of the star will depend on its orbital velocity and the kick onto the star from the supernova explosion. This distribution does not have a Gaussian profile, so we express the allowed positions as ellipses with a 99.73% probability (i.e., 3-σ) of containing the position of the ex-companion star. Since the proper motion depends on the nature of the companion, we report ellipses for red giants, subgiants, and main sequence stars. For SNR 0509-67.5 in particular, the shell expansion is uniform in all directions except for one quadrant where the interstellar medium is more dense (as shown by the excess 24-micron emission seen in the *Spitzer* image[27] from pre-existing dust swept up by the shell) and so the expansion has recently slowed down[28]. This slowing in only one quadrant accounts for the small observed ellipticity of the shell, from which we can derive the apparent offset (1.39"±0.14" along a line 18°±3° south of west) between the observed geometric center of the shell and the site of the supernova explosion. Our derived best estimate for the site of the explosion is J2000 05:09:30.976, -67:31:17.90. The error ellipse is nearly circular, with a conservative radius of 1.43" for a maximal proper motion (390 km/s), a maximal age for the remnant (550 years), and for 99.73% (3-σ) containment. (See Supplementary Information section 3 for details.)

The error ellipse is completely empty of all visible point sources down to the deep limits of *HST*. Importantly, there are no red giant or subgiant stars in or near the ellipse. (Red giants and subgiants can be confidently recognized by their position above the main sequence in the color-magnitude diagram.) The nearest red giant (star 'O' in Figure 1) is 7.4" from the center, while the nearest subgiant star (star 'N') is 5.8" from the center. The nearest star brighter than V=22.7 (star 'K'), i.e., the nearest possible ex-companion of any type, is 2.9" from the center. The only source in the ellipse is an extended faint nebula, and the excellent angular resolution of the *HST* allows us to see that no point source is hidden within the nebula. (This nebula is likely an irregular galaxy of moderate redshift, but the coincidence of this nebula with the site of the supernova is suggestive that its origin might be associated with the explosion, as discussed in Supplementary Information section 4.) The error ellipse is empty of point sources to a limiting magnitude of *V*=26.9 (at the 5-σ level). This requires that any ex-companion be less luminous than $M_V$=+8.4.

There is no red giant star in or near the error ellipse, and this is strongly inconsistent with the symbiotic progenitor model. There is no red giant or sub-giant star in or near the error ellipse, and this is strongly inconsistent with the recurrent nova, helium star, and spin-up/spin-down progenitor models. There is no star brighter than $V$=22.7 in or near the error ellipse, and this is strongly inconsistent with the supersoft source progenitor model. The lack of any possible ex-companion star to $M_V$=+8.4 rules out all published single-degenerate progenitor models. With all single-degenerate models eliminated, the only remaining progenitor model for SNR 0509-67.5 is the double-degenerate model.

**Supplementary Information** is linked to the online version of the paper at www.nature.com/nature.

**Acknowledgements** The *HST* images were taken as part of two programs with PI J. P. Hughes and PI K. Noll. This work was support with funds from the National Science Foundation.

**Author Contributions** B.E.S. and A.P shared in the ideas, the data analysis, and the writing of this paper.

**Author Information** The *HST* images are all in the public domain (see http://heritage.stsci.edu/2010/27/index.html, and http://apod.nasa.gov/apod/ap110125.html). Reprints and permissions information is available at www.nature.com/reprints. The authors declare no competing financial interests. Correspondence should be addressed to B.E.S. (schaefer@lsu.edu) or A.P. (pagnotta@phys.lsu.edu).

**Table 1.  Candidate Progenitor Classes**

| Candidate Class | $P_{orb}$ (days) | $V_{ex\text{-}comp}$ (km/s) | Surviving companion | $M_V$ (mag) | $V$ range in LMC (mag) |
|---|---|---|---|---|---|
| Double-degenerate | … | … | none | … | … |
| Recurrent nova | 0.6-520 | 50-350 | Red giant or subgiant | -2.5 to +3.5 | 16-22 |
| Symbiotic star | 245-5700 | 50-250 | Red giant | -2.5 to +0.5 | 16-19 |
| Supersoft source | 0.14-4.0 | 170-390 | Subgiant or >1.16 $M_o$ MS | +0.5 to +4.2 | 19-22.7 |
| Helium star donor | 0.04-160 | 50-350 | Red giant or subgiant core | -0.5 to +2.0 | 18-20.5 |
| Spin-Up/Spin-down | 245-5700 | 50-250 | Red giant or subgiant core | -0.5 to +2.0 | 18-20.5 |

The single-degenerate candidate classes mix together somewhat, with recurrent novae being temporary supersoft sources, some symbiotic systems appearing as supersoft sources, some recurrent novae being also technically symbiotic systems, and models showing that the long term evolution of a supersoft source is to become a recurrent nova before exploding as an SNe Ia.  The 'supersoft source' progenitor class is perhaps misnamed, as these nuclear burning white dwarfs can be emitting supersoft X-rays with large or small luminosity either episodically or persistently.  This table represents the basic classes proposed for progenitors, while other possibilities have been exhaustively examined and rejected (see Supplementary Information section 1).  The orbital periods, $P_{orb}$ in days, give an indication of the size of the companion star, and hence some indication of its brightness in isolation.  The velocity of the ex-companion star, $V_{ex\text{-}comp}$ in km/s, includes the orbital velocity of the companion with respect to the white dwarf plus the kick velocity, indicates the maximum proper motion that the companion star can have. The fourth column indicates the type of the ex-companion star for each candidate class, with 'MS' indicating a main sequence star. The absolute $V$-band magnitude ($M_V$) is that of the ex-companion star after the explosion. The visual magnitude ($V$) range for the ex-companion stars is for a distance modulus of 18.50 mag. A main point of this table is that the various classes of single-degenerate models all predict ex-companion star brightnesses that are 4.2 mag or more above our limit, and this is too large a gap to overcome by fading cores or stripped envelopes.

**Table 2.** Objects near the center of SNR 0509-67.5

| Star | RA & Declination (J2000) | $\Theta$ (") | $V$ (mag) | $I$ (mag) | Comments |
|---|---|---|---|---|---|
| A | 05:09:30.960 -67:31:16.28 | 1.7 | 26.08 ± 0.11 | 24.50 ± 0.08 | Nearest to error ellipse |
| B | 05:09:30.701 -67:31:18.75 | 1.7 | 24.82 ± 0.04 | 23.61 ± 0.04 | … |
| C | 05:09:30.753 -67:31:16.63 | 1.9 | 26.30 ± 0.13 | 24.77 ± 0.09 | … |
| D | 05:09:30.916 -67:31:19.91 | 2.0 | 24.02 ± 0.03 | 22.98 ± 0.03 | … |
| E | 05:09:30.660 -67:31:19.07 | 2.1 | 23.99 ± 0.02 | 23.05 ± 0.03 | … |
| F | 05:09:30.824 -67:31:16.03 | 2.1 | 23.30 ± 0.02 | 22.53 ± 0.02 | … |
| G | 05:09:31.212 -67:31:16.30 | 2.2 | 25.36 ± 0.06 | 23.76 ± 0.04 | … |
| H | 05:09:30.712 -67:31:16.01 | 2.5 | 22.87 ± 0.01 | 22.06 ± 0.02 | … |
| I | 05:09:30.581 -67:31:16.74 | 2.6 | 26.57 ± 0.15 | 24.72 ± 0.08 | … |
| J | 05:09:31.454 -67:31:17.21 | 2.9 | 25.84 ± 0.09 | 24.43 ± 0.07 | … |
| K | 05:09:30.824 -67:31:15.20 | 2.9 | 22.55 ± 0.01 | 21.86 ± 0.01 | Nearest $V$<22.7 |
| L | 05:09:31.299 -67:31:15.72 | 2.9 | 20.56 ± 0.01 | 20.07 ± 0.01 | … |
| M | 05:09:31.837 -67:31:19.61 | 5.2 | 24.26 ± 0.03 | 21.00 ± 0.01 | Very red star |
| N | 05:09:31.604 -67:31:22.54 | 5.8 | 20.92 ± 0.01 | 19.87 ± 0.01 | Nearest subgiant |
| O | 05:09:31.586 -67:31:11.49 | 7.4 | 18.75 ± 0.01 | 17.68 ± 0.01 | Nearest red giant |

The first column lists a letter name for each star for identification. The stars are labeled in Figure 1 with the letter placed to the immediate right of the star. The ordering is based on radial distance from the center of the error ellipse. The second column gives the position for each star. The third column gives the angular distance, $\Theta$, from the center of the error ellipse to the star. All stars with $\Theta$<3.0" are included, for the limiting magnitude of $V$=26.9 mag. Importantly, there are no stars within the extreme 99.73% error ellipse ($\Theta$<1.43"). Three additional stars of interest with $\Theta$>3.0" are added. The next two columns are the $V$ and $I$ magnitudes (with 1-$\sigma$ uncertainties), followed by a column for comments.

**FIGURE 1. SNR 0509-67.5 and the extreme 99.73% error ellipse.** The Hα image was taken with the *WFPC2* over three orbits in November 2007 with a total of 5000 seconds of exposure. The *B*, *V*, and *I* images were taken with the *WFC3* over two orbits in November 2010 with 1010, 696, and 800 seconds exposure respectively. North is up and east is to the left. These *HST* data were processed and combined with standard *PYRAF* and *IRAF* procedures. Figure 1 shows a combination of all four filters, with the remarkably smooth Hα shell visible. The error circle (with 1.43" radius) is the extreme 99.73% region (3-σ), where to be on the edge the ex-companion star must be a main sequence star with the minimum possible mass for any published model (1.16 $M_0$), the velocity must be entirely perpendicular to the line of sight, the age of the supernova remnant must be pushed to the 3-σ highest possible value (550 years), and the measurement error for the remnant's geometric center must be pushed to the 3-σ extreme. The only source inside the error ellipse is a nebulous object that looks like a background galaxy, however the location of this object at the center suggests it might be related to the supernova event (see Supplementary Information section 4). There are no stars within the extreme error circle to *V*=26.9 mag, which corresponds to an absolute magnitude of $M_V$=+8.4 mag in the LMC. All published models for single-degenerate progenitors have the ex-companion star appearing more luminous than $M_V$=+4.2 (*V*=22.7 in the LMC). In all, our extreme 99.73% error circle is very conservative, and there is no point source to limits 4.2 mag deeper than possible for any published model of single-degenerate systems.

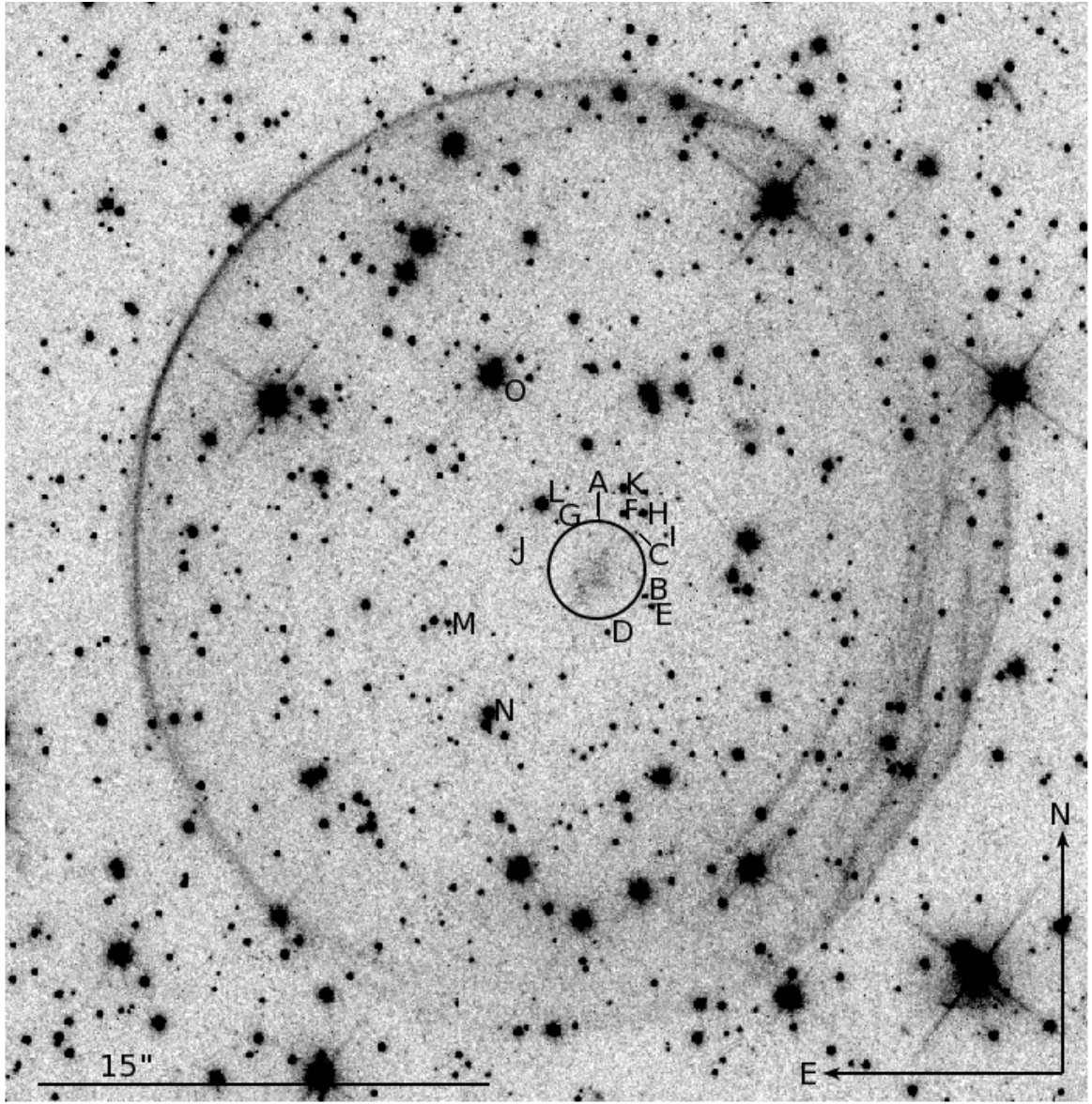

**Supplementary Information**

1 Are there any viable models involving low-luminosity companions?

2 Geometric center of SNR 0509-67.5

3 Expected offset of the ex-companion star from the geometric center of the remnant

4 The nebula in the middle of the error ellipse

**Supplementary Table S1** 99.73% error ellipses for SNR 0509-67.5

# Supplementary Information

**1. Are there any viable progenitor models involving low-luminosity companions?**

Our primary observational result is that the center region of SNR 0509-67.5 is empty to $V$=26.9 mag, with our primary analysis concluding that this precludes any single-degenerate progenitor. Yet, is it possible that some viable single-degenerate model can have an ex-companion star that will appear fainter than $V$=26.9 mag?

Previously, weak support has been given to the short orbital period (1.92 hours) recurrent nova T Pyx as a candidate progenitor[29,30], and the companion star for such a short orbital period has a low luminosity and a mass of 0.14±0.03 $M_o$[31]. But the high accretion rate of T Pyx is due to irradiation of the companion from a previous ordinary nova event[30,32], and this is declining greatly on the time scale of a century, so the recurrent nova episode of high accretion is only a short interval out of a much longer quiescent interval[32]. The short interval of recurrent nova events might or might not lead to an increase in the mass of the white dwarf, but the ordinary nova event will completely dominate the few recurrent nova events and expel more matter from the white dwarf than is accreted over the whole cycle[33], so the T Pyx white dwarf is losing mass and will not become an SN Ia.

Persistent supersoft sources are reasonable progenitor candidates and they can have main sequence companion stars, so we must determine what is the faintest possible such star. These systems "are binaries containing white dwarfs which can accrete matter from a more massive and possibly slightly evolved companion."[34] Orbital periods range from 0.14 to 3.5 days, with the shorter period systems having too little mass to allow the white dwarf to reach the Chandrasekhar limit. The fast accretion onto the white dwarf (which is required to power the steady hydrogen burning that produces the persistent supersoft X-ray light) is driven by the Roche lobe shrinking faster than the companion star[35], which requires[36] a mass ratio of >5/6. For the white dwarf to be near the Chandrasekhar limit, this requires that the companion star be more massive than 1.16 $M_o$. Such a star will necessarily be at least as luminous as a normal 1.16 $M_o$ main sequence star, for which the absolute magnitude is $M_V$=+4.2. This result has confirmed by very detailed models[37]. With the LMC distance modulus of 18.50, the star would appear brighter than $V$=22.7 mag.

Progenitor models have been proposed wherein the companion star has been stripped of most of its outer envelope, so we should consider whether these can produce low-luminosity ex-companion stars. One such model is that of helium star companions, red giants stripped of their outer hydrogen envelope, with the remaining helium envelope providing the mass accreted onto the white dwarf. But the donor star still has the same energy generation as in the core of the original red giant, so the luminosity is still 1000 to 10,000 times that of the Sun and at temperatures around 80,000° K[38]. This compact star will suffer relatively little mass loss during the supernova explosion[23]. The absolute magnitude of such a star will be roughly $M_V$=+2 or brighter (including bolometric corrections), so any such ex-companion in the LMC will appear as $V$=20.5 mag or brighter.

The spin-up/spin-down model[39,40] posits a red giant donor star that spins up the white dwarf so that its rotation will support a mass greatly exceeding the Chandrasekhar limit, until the donor's envelope is exhausted and the donor star shrinks to a small, hot core, while the white dwarf takes a longer time to redistribute or lose angular momentum to allow for the ignition of the supernova event. (The published model is for a red giant or possibly a subgiant companion[39,40], but in principle this could be extended to main sequence stars[40]. The name 'spin-up/spin-down' refers to the progenitor model, but it can also refer to the physical process where the white dwarf spins up and then spins down. The spin-up process is inevitable and previously ignored, although the spin-down process will only occur in this model in the small-chance case that the companion star turns off the accretion when the white dwarf is above the Chandrasekhar mass.) The result will be a relatively small ex-companion star with little surface material blown off by the supernova[23]. Again, the core of the red giant star will have the same luminosity as before the explosion. The time from the cessation of the accretion (after which the companion's exhausted envelope shrinks on the fast Kelvin-Helmholtz time scale) until the supernova event occurs is governed by the growth rate for r-mode instabilities that will redistribute or lose angular momentum from the white dwarf. Calculations of this growth rate[41] for the relevant conditions gives time scales of $10^3$ to $10^5$ years[42]. During this time, the luminosity of the companion will change little, so that a typical luminosity is 50 solar luminosities, which for the given temperature of 6000° K corresponds to $V$=19.0 mag in the LMC[40]. (The delay would have to be roughly $10^9$ years for the ex-companion to cool and fade below our limit of $V$=26.9 mag[40].) So any ex-companion from a spin-up/spin-down progenitor in the LMC must appear brightly near the middle of our error ellipse.

We can also consider an idea within a spin-up/spin-down scenario where the companion is a main sequence star that might somehow get to low luminosity before the explosion. At the start of this scenario, the only means for the white dwarf to spin up and gain mass is for the accretion rate to be very high, which can only be when the mass ratio is >5/6. Then, as the mass of the main sequence companion falls below 1.16 M$_o$ (with $M_V$=+4.2), the accretion rate will largely turn-off. The hallmark of the spin-up/spin-down idea is that the delay from the end of spin-up to the explosion allows for the companion to shrink (so as to minimize the hydrogen contamination of the subsequent supernova shell as well as to minimize the Kasen effect[43]). But the 1.16 M$_o$ star will be unchanging on any interesting time scale. The system will still have a relatively low accretion rate (driven by angular momentum loss due to magnetic breaking) that will very slowly reduce the mass of the companion from 1.16 M$_o$ (with $M_V$=+4.2) down to ~0.5 M$_o$ (with $M_V$=+8.4). The time required for ordinary magnetic breaking to grind down the companion star is roughly $5 \times 10^9$ years[44]. Indeed, the time scale for the companion star to start evolving off the main sequence is likely faster, in which case its luminosity will be *brightening*. In all cases, the time it takes for a main sequence companion star to diminish to the point where it would be invisible inside the SNR 0509-67.5 error ellipse ($5 \times 10^9$ years) is many orders of magnitude longer than the delay time between the end of the spin-up and the explosion ($10^3$ to $10^5$ years[42]). In all, the spin-up/spin-down model with a main sequence star cannot produce a low luminosity ex-companion star because the companion star will be at $M_V$=+4.2 when the fast accretion stops and it will still be at $M_V$=+4.2 when the explosion happens.

Exhaustive analyses of all combinations of star models and observed binary systems with white dwarfs have examined various types of single-degenerate systems that could conceivably produce SNe Ia[1,45]. After this consideration, all the reasonable single-degenerate systems either had evolved luminous companions or main sequence companions with more than one solar mass. Systems with low-mass main sequence stars (the cataclysmic variables) were rejected both because they could not maintain the high required accretion rate necessary to avoid hydrogen flashes (which makes the white dwarf lose mass over the long term) and because the number density and death rate of these systems are greatly too low to account for the observed rate of SNe Ia[1].

We should also consider the possibility that the supernova explosion itself could modify and dim the companion star significantly. For the cases where the companion star has a moderate or high surface gravity (the main sequence stars in supersoft progenitors, helium donor stars, and the cores in spin-up/spin-down progenitors), the stripping of the envelope will be minimal and the pre-explosion star will have much the same luminosity as 400 years after the explosion[22-24]. Detailed calculations for the subgiant case show that usually the ex-companion star will be up to two orders of magnitude more luminous (due to the deposited energy), although in the unexpected case of low energy deposition the ex-companion can be as much as ten times less luminous (due to internal energy going into expanding the surviving envelope)[24]. In all these cases, the stellar core is still producing energy at the same rates, so the luminosity cannot change greatly. Even with an unexpected dimming of a factor of ten (2.5 magnitudes), the ex-companion stars for all proposed progenitor classes will still be more than a factor of ten brighter than our deep limits for SNR 0509-67.5.

The bottom line is that there are no published single-degenerate models for which the ex-companion star will be significantly less luminous than $M_V$=+4.2 ($V$=22.7 mag in the LMC).

## 2. Geometric center of SNR 0509-67.5

Any ex-companion star should appear near the geometric center of the shell. The shell of SNR 0509-67.5 is nearly symmetric and smooth, making this a good case for measuring an accurate center position. But the shell center cannot be measured perfectly, and different measures will yield different centers. Here, we report on three independent methods to determine the geometric center. Importantly, these methods use different gases in different positions of the shell.

The first method defines the center based on the outer edge of the Hα shell. The procedure is to take a baseline cut through the shell, noting the very edges, taking the perpendicular bisector of this segment, noting the very edges, and taking the center to be the bisector of this perpendicular segment. A total of nine such centers are obtained for baselines tilted at 10° intervals (see Supplementary Table S1), to sample the entire edge of the shell. The nine centers are then averaged to get a combined center, and the RMS scatter of these nine positions is a measure of the 1-σ accuracy of this combined center position. For the nine tilted baselines, the table specifies the offsets from the combined center in terms of right ascension (ΔRA) and declination (Δδ) as expressed in arc-seconds. In practice, this procedure is iterated once so as to avoid any sensitivity to the initial assumed center. All 36 measured edge positions define the shell radius as a function of angle from north. This radius function is closely a sine wave, except for the deviation associated with the moderately extended wispy filament towards the northwest edge of the shell. A $\chi^2$ fit gives a radius of 16.0" along the long axis (oriented to 18°±3° west of north) and a radius of 14.6" along the short axis. The ratio of the short axis to the long axis is 0.913±0.009. The error ellipses are quoted in the direction of these long and short axes. The center and uncertainties from this first method are presented in Supplementary Table S1.

The second method defines the center based on the outer edge of the X-ray shell. For this, we have used three *Chandra* images[46] from May 2000, in which the remnant was imaged separately in the light of three emission lines: O (0.45-0.7 keV), Fe L (0.7-1.4 keV), and Si (1.5-2 keV). The procedure for finding the center of the three X-ray images is the same as the first method, with the three resultant centers being closely consistent and averaged together to get one combined center based on the edge of the X-ray shell. The uncertainty in this position is characterized by the RMS scatter (in the direction of the long and short axes) of the individual centers. Supplementary Table S1 gives this position and error ellipse.

The third method uses the faint Hα light in the remnant's central region. This interior light, far inside the outer filaments, is visibly faintest near the geometric center. This is simply the thin shell seen nearly perpendicular to its surface (instead of being seen edge-on near the edge of the remnant, which creates the thin filaments). For a thin shell of radius $R_{shell}$, the brightness falls off with distance $R$ from the remnant's center as $I_{back}+I_{center}*(1-[R/R_{shell}]^2)^{-0.5}$, with $I_{center}$ being the central brightness. We measure the brightness of 20x20 pixel tiles in the interior of the shell, and then fit them to this brightness model. (We have also made model fits where $R_{shell}$ is allowed to vary as an ellipse, with essentially identical resulting centers.) We used 71 tiles within 110 pixels of the center (iteratively determined) for which the maximum pixel value in the tile was <0.001 counts per second. The uncertainty on each tile brightness was taken to be the RMS scatter of tiles outside the remnant, while the average for these tiles was taken to be the background brightness $I_{back}$. We use a $\chi^2$ fit to determine the best center, and the 1-σ error bars along the long and short axes are determined by the point at which the $\chi^2$ value has risen by unity above its minimum. Our best fit model has a $\chi^2$ of 61.3 (for 67 degrees of freedom). We get the same results (to within the 1-σ error bar) if we use different tile sizes, different radial cutoffs, and different star rejection thresholds. Our best fit center and the 1-σ errors along the two axes are presented in Supplementary Table S1.

We now have three independent geometric centers for the shell; each measure is based on a different gas or region. The first method is based on the relatively cold gas around the visible edge, the second method is based on the very hot gas around the edge, and the third method is based on the relatively cold gas near the middle. We have combined these three independent positions as a weighted average. Our final result for the geometric center of the shell is J2000 05:09:31.208, -67:31:17.48 with 1-σ uncertainties of 0.14" and 0.20" in the short and long axes respectively (see Supplementary Table S1).

# 3. Expected offset of the ex-companion star from the geometric center of the remnant

Any ex-companion star is unlikely to appear at the exact geometric center of the remnant for several reasons, including the proper motion of the star away from the site of the explosion, the possibly asymmetric ejection of material so that the geometric center of the observed shell is offset from the site of the explosion, and the possibly asymmetric distribution of gas in the interstellar medium that slows the shell expansion in some direction more than in the opposite direction, resulting in an offset between the observed geometric center of the shell and the site of the explosion. Explosion sites have not been directly measured for any SN Ia, so we must evaluate the expected sizes of these offsets from the simple physics of the situation. (There are extensive measures of the offsets of neutron stars from core collapse supernovae[47], but the physical setting is greatly different from the SN Ia case, so this experience has no utility for understanding the offset of our LMC remnant.)

The proper motion of the ex-companion star (with respect to the center of mass of the original binary system) will come from both the kick given to the star by the supernova ejecta and the orbital velocity at the time of the explosion. The kicks onto the companion from the supernova ejecta will always be relatively small[13,22,23]. For companions filling their Roche lobe, the orbital velocity will depend primarily on the stellar radius. Canal et al.[13] calculated average post-explosion velocities for expected conditions, with the conclusion that the ex-companions should be moving at around 480, 250, and 100 km/s for main sequence, subgiant, and red giant companions, respectively. For the red giant and subgiant cases, the proper motion is relatively small and all such stars are far outside the error ellipses. The only critical case is when we push to the smallest possible mass main sequence star, which produces the largest possible error ellipse (see Figure 1). The smallest mass main sequence star that can be a companion star for an SN Ia is a 1.16 $M_o$ star in a supersoft system (see Supplementary Information section 1). The 480 km/s velocity from Canal et al. is for a 0.6 $M_o$ star, and the proper motion gets smaller as the companion mass increases. For a 1.16 $M_o$ main sequence companion star filling its Roche lobe around a 1.4 $M_o$ white dwarf, the orbital period will be 10.6 hours, the orbital velocity of the companion star will be 208 km/s, and the white dwarf orbital velocity will be 173 km/s. The supernova explosion will provide a kick to the companion star of 86 km/s in the direction perpendicular to the orbital motion[22]. The relative velocity of the white dwarf (which will be the origin for the frame of the expanding shell) and the companion star will be 390 km/s. Going to higher mass main sequence stars will only make for a smaller velocity. So for all viable progenitor models, the velocity of the ex-companion with respect to the original geometric center of the remnant will be 390 km/s or less. For an LMC distance modulus of 18.50±0.10, the extreme case (390 km/s in a tangential direction) results in a total proper motion of 0.0016 "/year. For the 400±50 year age of SNR 0509-67.5, any ex-companion star must be within 0.66"±0.08" of the site of the explosion.

        Largely, the thermonuclear burning of the white dwarf is spherically symmetric, so any asymmetries will be small. Observationally, asymmetries can be measured by polarization studies, where normal SNe Ia have small polarization in the spectral continuum (up to 0.2%-0.3%), which is consistent with an ellipsoidal shape where the minor-to-major axis ratio is 0.9[48-50]. The asphericity might be smaller if the polarization is caused by dense clumps occulting part of the photosphere[51]. The observed axis ratio for SNR 0509-67.5 is 0.913±0.009. If this asphericity is dipolar in shape (e.g., oblate or prolate), then the geometric center of the shell will correspond to the original position of the binary. The shell center will be offset only if there is some appreciable monopolar component (e.g., where the north pole is ejected with higher velocity than the south pole). Even for monopolar asymmetries, the apparent offset will generally be smaller than the maximal value due to projection effects, and such offsets will be near zero for cases where the monopolar axis is near the line of sight. In theory, an off-center detonation in the white dwarf might result in asymmetric distributions of density and composition, and this will create apparent velocity differences (as viewed from opposite directions) as the photosphere recedes at differing rates[52]. This scenario is apparently confirmed[52] by strong correlation of the velocity gradients (with high and low groups) and the bulk velocities at late times (with redshifted and blueshifted groups), as well as by the lopsided distribution of opacity in the sub-luminous SN Ia S And[53]. The model predicts late-time velocity differences (between hemispheres) of less than 10%, but this is mainly an effect of different photospheric depths, and it is unclear whether the off-center detonation

translates into an offset of the geometric center of the shell. From these considerations, the maximum offset of the geometric center of the shell from the original explosion position is roughly 10% of the radius.

A global gradient in the density of the interstellar medium across the shell will result in the remnant having different radii in different directions, causing an apparent offset of the geometrical center from the site of the original explosion. SNe Ia are generally in low density environments, so this effect is likely to be small. Indeed, *Spitzer* observations show no significant background flux around SNR 0509-67.5[54], while extinction maps show no significant gradients across the remnants[55]. Badenes et al.[56] characterizes SNR 0509-67.5 as being "in a very homogenous region".

A measure of both asymmetry offsets can be obtained from the observed ellipticity of the shell. In the case of either a lopsided high ejecta velocity or a low interstellar medium density in some direction, the out-of-round shape is due to the shell having a large radius in that direction (`$f$` times the radius in other directions, with $f>1$). In this case, the observed short-to-long axial ratio will be $2/(1+f)$, while the offset between the site of the explosion and the observed shell center will be $0.5(f-1)R_{shell}$ in one direction or the other along the long axis. In the case of either a lopsided low ejecta velocity or a high interstellar medium density in some direction, the out-of-round shape is due to the shell having a small radius in that direction (with $f<1$). With this, the observed short-to-long axial ratio will be $(1+f)/2$, while the offset will be $0.5(1-f)R_{shell}$ in one direction or the other along the short axis. In all four cases (high/low ejection velocity or high/low interstellar medium density in some direction), if the direction is not perpendicular to the line of sight, then the foreshortening of the offset will be evident in the reduction in the ellipticity of the shell.

The case of SN1006 provides a beautiful example of how our method recovers the site of the original explosion. This thousand year old galactic remnant is nicely symmetrical, with a small ellipticity. The long axis is along the NNE-SSW line and the ratio of the short axis to the long axis is 0.90. From this, we get $f=1.22$ and a fractional offset of 11%. If we only had the shape of the shell, we would not know the direction of this 11% offset between the geometric center and the site of the explosion. (High ejecta velocity or low ISM density in one quadrant will result in an offset that is 11% either towards the NNE or the SSW, while low ejecta velocity or high ISM density in one quadrant will result in an offset that is 11% towards the ESE or WNW.) For SN 1006, this ambiguity can be resolved using absorption spectroscopy of five background sources, where the results show that the supernova ejected high velocity material towards the NNE quadrant[57]. In the three-dimensional analysis, the geometric center is offset by roughly 20% of the shell radius, although when projected onto the sky this corresponds to an offset of only roughly 10% of the shell's angular radius. With this, the direction ambiguity is resolved such that the offset from the observed geometric center to the explosion site is 11% towards the SSE. The good agreement between the offset from our analysis (based on the observed ellipticity of the shell) and the full three-dimensional analysis is heartening. However, we see that we must have a means to break the direction ambiguity, as otherwise we have a substantially larger error ellipse.

For the case of SNR 0509-67.5, we can cleanly choose between the four alternative offset possibilities and determine the offset and direction. The *Spitzer* 24-micron image shows the pre-existing and swept-up dust from the surrounding interstellar medium, and there is an excess of swept-up material in the quadrant centered towards the WSW short axis[58]. The swept-up material towards the NNW, the ENE, and the SSE axes is identical (as seen in the dust brightnesses in those directions), and is significantly lower than the amount in the WSW direction. This explains why the short axis of the shell is in that direction. This is supported[28] by an analysis of the X-ray line widths where the shock velocity in the SW quadrant (5000 km/s) is somewhat lower than for the NE quadrant (6000 km/s), with the slow-down towards the SW being relatively recent. So the case of high interstellar density in one direction (to the WSW) is known, and this results in an offset from the geometric center to the explosion site towards the WSW. For an axial ratio of 0.913±0.009, we have $f$=0.826±0.018 and an offset of 1.39"±0.14". The uncertainty in the direction of the short axis (±3°) makes for an uncertainty of the offset position of 0.07" in the direction of the long axis of the shell. With this offset and its added uncertainty, our measured position for the site of the supernova event is J2000 05:09:30.976, -67:31:17.90 with 1-σ uncertainties of 0.21" and 0.20" in the long and short axes respectively.

The true difference between the observed geometric center and the position of the ex-companion star will arise from the proper motion of the ex-companion (relative to the white dwarf and including the kick from the supernova), the uncertainty in measuring the geometric center of the shell, and the offset of the geometric center from the site of the supernova due to the relatively high density of the interstellar medium towards the WSW. A complication arises because the distribution of the offsets from proper motion is not Gaussian shaped (rather, it is edge dominated), so the size of the ellipse for the position of the ex-companion star cannot be simply expressed with a Gaussian sigma. To account for this, we have constructed Monte Carlo simulations of the various mechanisms, including the random orientation of the proper motion, the random error in the age of the supernova remnant, and the Gaussian random error in measuring the geometric center of the shell. We report the long and short radii for ellipses (oriented with the axes in the same direction as the shell) such that 99.73% (i.e., 3-$\sigma$) of the realizations are within the ellipse. Since the possible proper motions have a circularly symmetric distribution and the position for the site of the supernova event has nearly identical uncertainties in the two axes, we can make an accurate simplification that the final error ellipses are error circles. These error circles will depend on the probability level for containing the ex-companion (e.g., 99.73%) and on the adopted proper motion of the ex-companion star (typically 100 km/s for red giants, 250 km/s for subgiants, and 390 km/s for main sequence stars). The 99.73% error circle radii are 0.74" for red giants, 1.06" for subgiants, and 1.17" for main sequence stars. For the most conservative case (a 1.16 $M_0$ main sequence companion and a 550 year old remnant), the 3-$\sigma$ error circle is 1.43" in radius. Thus, our main result is that any ex-companion star of SNR 0509-67.5 must be within 1.43" of J2000 05:09:30.976, -67:31:17.90 (see Figure 1).

      A combination of fortuitous circumstances allows for our small error circle (roughly 10% of the shell radius). First, the supernova is quite young (400±50 years), so the companion star has not had much time to move far from the site of the explosion. Second, the shell is nicely symmetrical, and this allows us to accurately determine the geometric center. Third, the *Spitzer* images demonstrate that the shell's ellipticity is caused by a somewhat denser interstellar medium in one quadrant, which resolves the direction of the offset. In all, the maximum radius of our error circle is 1.43". The ex-companion can lie at this extreme only for the case where it is a main sequence star, it has

the lowest acceptable mass (1.16 $M_0$), the age of the remnant is pushed to its 3-σ high value (550 years), the velocity of the companion star is entirely perpendicular to the line of sight, and the measurement errors on the geometric center are at their 3-σ extreme. Without such extreme assumptions all occurring together, a main sequence ex-companion has a two-thirds chance of being in the innermost 0.7" of our error circle.

## 4. The nebula in the middle of the error ellipse

The center of our error ellipse contains a nebula that might or might not be a background galaxy. The integrated magnitude for the nebula is $V=23.32\pm0.07$ and $I=20.95\pm0.02$, with a red color. The nebula appears faint in the H$\alpha$ image, so this is not simply some shard of the outer shell. This nebula has an extended area roughly 2.1"x1.4", with a central bright core plus 3-6 knots within this contiguous area, as well as ~6 isolated, faint, and extended knots outside the main nebula. The center of this nebula is 0.2" from our best estimate of the position of the supernova explosion. The contiguous region has a maximal distance from the central core of 1.3", while the farthest of the isolated knots is 2.0" from the center.

There can be no point source hidden by this nebulosity to the stated limit of $V=26.9$ mag. To give specific numbers, the V-band image has the brightness in the brightest 3X3 pixel box for the brightest knot equal to 0.15 e/pixel/sec above the background, whereas star A ($V=26.08$, see Table 2) has its brightest 3X3 box equal to 0.33 e/pixel/second above background, which puts the brightest knot at $V=26.9$. All the knots are definitely extended. No significant source with a point spread function rises above the nebula.

The obvious idea is that this nebula is a background galaxy of no relevance to the supernova. The mottled shape and color are like other galaxies at moderate redshift as seen by *HST*. This is reinforced by the presence of four other similarly red and extended galaxies just outside the supernova shell.

Nevertheless, this nebula is strikingly centered at the site of the explosion, and this is suggestive of a connection. With five such objects (red and extended) in the 4500 square arc-second field of view, the probability of a red nebula appearing inside our 1.60" radius error circle (with area 8.0 square arc-seconds) is 0.9%, although such *a posteriori* calculations are always problematic. If the nebula is associated with the supernova, then this might represent very low velocity ejecta left far behind by all the other ejected mass. An alternative idea is that the nebula comes from a double-degenerate progenitor system where the low mass white dwarf would form a temporary accretion disk as it disrupts when the high mass white dwarf explodes before all of the material can be accreted[59], so the remaining accretion disk material would fly away at typical orbital velocities. For the observed nebula, the size and age yields a characteristic velocity for the contiguous region equal to 800 km/s, while the farthest isolated knot would have a velocity of 1200 km/s or more. We know of no precedent for such low-velocity material. A possible way to distinguish the likely galaxy identity from the ejecta possibility is to get a spectrum of the nebula, where any ejecta should be bright in emission lines.

## Supplementary Table S1

**99.73% error ellipses for SNR 0509-67.5**

| Position | Center RA & Dec (J2000) | | $\sigma_{short}$ | $\sigma_{long}$ |
|---|---|---|---|---|
| Hα center for 0° cross ($\Delta RA$=-0.08", $\Delta\delta$=0.00") | 05:09:31.159 | -67:31:17.17 | … | … |
| Hα center for 10° cross ($\Delta RA$=0.01", $\Delta\delta$=-0.21") | 05:09:31.143 | -67:31:17.38 | … | … |
| Hα center for 20° cross ($\Delta RA$=0.09", $\Delta\delta$=-0.22") | 05:09:31.128 | -67:31:17.39 | … | … |
| Hα center for 30° cross ($\Delta RA$=0.27", $\Delta\delta$=0.08") | 05:09:31.098 | -67:31:17.09 | … | … |
| Hα center for 40° cross ($\Delta RA$=0.31", $\Delta\delta$=0.38") | 05:09:31.091 | -67:31:16.79 | … | … |
| Hα center for 50° cross ($\Delta RA$=0.27", $\Delta\delta$=0.66") | 05:09:31.098 | -67:31:16.51 | … | … |
| Hα center for 60° cross ($\Delta RA$=-0.24", $\Delta\delta$=-0.19") | 05:09:31.187 | -67:31:17.36 | … | … |
| Hα center for 70° cross ($\Delta RA$=-0.26", $\Delta\delta$=-0.19") | 05:09:31.190 | -67:31:17.36 | … | … |
| Hα center for 80° cross ($\Delta RA$=-0.35", $\Delta\delta$=-0.27") | 05:09:31.206 | -67:31:17.44 | … | … |
| Combined center of Hα edge (method 1) | 05:09:31.144 | -67:31:17.17 | 0.18" | 0.37" |
| Center of X-ray edge (method 2) | 05:09:31.195 | -67:31:17.11 | 0.26" | 0.26" |
| Minimum of Hα interior light (method 3) | 05:09:31.342 | -67:31:18.34 | 0.54" | 0.60" |
| Geometric center of shell (methods 1-3) | 05:09:31.208 | -67:31:17.48 | 0.14" | 0.20" |
| Site of supernova explosion | 05:09:30.976 | -67:31:17.90 | 0.20" | 0.21" |
| Ex-companion star, red giant proper motion | 05:09:30.976 | -67:31:17.90 | 0.74" | 0.74" |
| Ex-companion star, subgiant proper motion | 05:09:30.976 | -67:31:17.90 | 1.06" | 1.06" |
| Ex-companion star, main sequence proper motion | 05:09:30.976 | -67:31:17.90 | 1.17" | 1.17" |
| Ex-companion star, extreme 99.73% error circle | 05:09:30.976 | -67:31:17.90 | 1.43" | 1.43" |

The columns for $\sigma_{long}$ and $\sigma_{short}$ give the uncertainties in the long axis (18°±3° west of north) and the perpendicular short axis. The quoted uncertainties for the positions of the ex-companion star are all for the 99.73% (3-σ) probability level, while the other quoted uncertainties are for the usual 1-σ probability level.